\documentclass{scrartcl}
 \usepackage{graphicx}    
\usepackage[normal]{caption}

\usepackage{graphicx}
\usepackage{amsmath} 

\begin{document}
\title{Luminosity Measurement at COMPASS}
\author{C. H\"oppner\footnote{Technische
  Universit\"at M\"unchen, Physik Department, 85748 Garching,
  Germany} \footnote{ email: christian.hoeppner@cern.ch,
  supported by the German Bundesministerium f\"ur Bildung und Forschung
  and the DFG cluster of excellence ``Origin and Structure of the Universe''.}
  for the COMPASS Collaboration}
\date{}
\maketitle
\abstract{The measurement of
    absolutely normalized cross sections for high-energy scattering
    processes is an important reference for theoretical models. This
    paper discusses the first determination of the luminosity for data
    of the COMPASS experiment, which is the basis for such measurements. The resulting normalization is validated via the determination of the structure function $F_2$ from COMPASS data, which is compared to literature.} 

\section{Introduction}
\label{sec:1}
COMPASS \cite{COMPASS} is a fixed-target experiment at the CERN SPS
which investigates the spin structure of nucleons and hadron
spectroscopy with high-intensity muon and hadron
beams. Nucleon spin-structure measurements are performed by scattering
polarized muons off polarized solid-state targets ($^6$LiD or NH$_3$).
COMPASS has published several important results in this field, for
instance on gluon polarization from hadron-pair production with high
transverse momenta (high-$p_T$) \cite{COMPASShighpt} and from
open-charm production \cite{COMPASSopencharm}, on the spin-dependent
structure function $g_1$ \cite{COMPASSg1p}, on Collins and Sivers
asymmetries \cite{COMPASStrans}, or on quark helicity distributions in
nucleons \cite{COMPASShel}. These results have been obtained from
direct measurements of double-spin asymmetries in which the absolute
luminosity normalization cancels.  Measurements of absolute
interaction cross sections were not yet performed because the
experiment does not have a dedicated luminosity monitor.
However, such measurements provide important
benchmarks for the ability of theoretical models to describe 
experimental data from high energy physics experiments.
Examples are cross sections for high-$p_T$ particle production,
open-charm production, or exclusive photon production in the future
Deeply Virtual Compton Scattering (DVCS) program of COMPASS-II
\cite{COMPASS2}.  The analysis presented here was performed in the
framework of the measurement of the cross section for quasi-real
photo-production of charged hadrons with high $p_T$ in muon-deuteron
scattering. The unpolarized and polarized cross sections for this
process have been calculated in next-to-leading order perturbative
quantum chromodynamics (NLO pQCD) \cite{Jaeger}.  A comparison of the
calculated polarized cross section with the experimental spin
asymmetries can be used to constrain the gluon-polarization
distribution $\Delta g/g(x_g)$, which is an input for the
calculation. Before this extraction can be carried out with
confidence, the theory first has to be capable of correctly predicting
the unpolarized cross section, which will soon be published by the
COMPASS collaboration.

This paper discusses the underlying estimation of the luminosity for the
COMPASS data and presents a validation of the result via the
measurement of the well known cross section for inclusive muon
scattering. The paper is structured as
follows: The terminology of cross section and luminosity measurements
in the context of COMPASS is introduced in Sec.\ \ref{sec:2}. The
used data set is described in Sec.\ \ref{sec:3}, followed by the data
selection criteria in Sec.\ \ref{sec:4}. Section \ref{sec:5}
describes the measurement of the beam flux and is followed by
Sec. \ref{sec:6} which explains the sources of dead times in the
measurement and their corrections. Section \ref{sec:7} discusses the
determination of the structure function $F_2$ using the resulting
integrated luminosity. The comparison of the results to a
parametrization obtained from measurements of $F_2$ by the NMC
experiment \cite{NMC} confirms that the COMPASS luminosity has been
correctly determined within a systematic uncertainty of 10\%. The
normalized data set can hence be used to measure new and unknown
unpolarized cross sections.
\section{Luminosity}
\label{sec:2}
The interaction cross section for the observation of a particular final state is defined as:
\begin{align}
\sigma = \frac{\dot{N}}{\mathcal{L}} = \frac{N}{L}
\end{align}
with the rate of occurrence of the final state $\dot{N}$ and the
instantaneous luminosity $\mathcal{L}$ (and their respective time integrals
$N$ and $L$).  For fixed target experiments, the instantaneous luminosity is
defined as
\begin{align}\label{eq:beam}
\mathcal{L} = \Phi_{\text{beam}} \cdot N_{\text{target}}=
\frac{R_{\text{beam}}}{A_{\text{target}}}\cdot N_{\text{target}}
\end{align}
where $\Phi_{\text{beam}}$ ($R_{\text{beam}}$ ) is the beam flux
(rate) through the fiducial target volume, $A_{\text{target}}$ is the
area of the fiducial target, and $N_{\text{target}}$ is the number of
target nucleons in the fiducial target volume. The fiducial target
volume is defined as the part of the target volume which is retained
after the geometrical cuts on the primary
vertices. Only event with beam tracks which cross the full length of the fiducial target volume are used for the analysis.\\
The observation of final states can be affected by misreconstruction
of kinematical variables, detection inefficiencies, and dead times in
which the experiment can not record events.  The kinematical smearing
and the detection inefficiencies, which are mostly due to incomplete
geometrical coverage of the phase space by the detectors and trigger
elements, are summarized in the acceptance correction factor
$\epsilon$. The cross section is then given as
\begin{align}\label{eq:xs}
\sigma = \frac{\tilde{N}/\epsilon}{\tilde{L}}
\end{align}
with the number of \emph{observed} final states $\tilde{N}$ and the effective
integrated luminosity $\tilde{L}$, which is corrected for the dead
times of the experiment.\\
The COMPASS beam is delivered by the SPS accelerator in so-called
spills\footnote{ the term \emph{spill} denotes an extraction from the accelerator.}. In the 2004 beam time, when the discussed data set was
taken, COMPASS was supplied with spills of muon beam of length 4.8\,s,
followed by breaks of 12\,s.  The dead times in the data taking caused
by the data acquisition system (DAQ) and the veto system of the
scattered-muon trigger \cite{triggerNIM} are rate dependent.  Since the
intensity of different spills can vary considerably, the dead times
need to be corrected on a spill-by-spill basis. The acceptance
correction factor $\epsilon$, which is obtained from a Monte Carlo
simulation of the experiment with a constant beam rate assumption,
only includes effects which are not or only weakly rate dependent. All
rate-dependent effects are absorbed into the definition of the
effective integrated luminosity of a spill $i$
\begin{align}\label{eq:effLumi}
\tilde{L}_i = \int_{\text{time in spill }i}
[\mathcal{L}_i(1-d_{i, \text{DAQ}})(1-d_{i, \text{veto}})] dt.
\end{align}
where $d_{i, \text{DAQ}}$ is the  DAQ dead time, i.e.\ the fraction of
data taking time in which the DAQ can not accept triggers because it
is busy with readout of previously triggered events, and $d_{i,
  \text{veto}}$ is the dead time of the veto system of the muon
trigger, i.e.\ the duty cycle of the veto signal (for details see
Sec.\ \ref{sec:6}).
The total integrated luminosity $\tilde{L}$ is the sum over all spills
which are used for the extraction of the number of final states $\tilde{N}$.

\section{Data Set}
\label{sec:3}
The analyzed data set was recorded in 2004. This particular choice is
motivated by the fact that the measurement of the hadron production
cross section depends on semi-inclusive trigger systems which include
the response of the two hadronic calorimeters of COMPASS. After the
introduction of a new electromagnetic calorimeter (ECAL1) after the
2004 beam time, the efficiencies of these triggers have been
compromised, which would make an absolutely normalized measurement
more difficult, while spin asymmetry measurements are not affected. In
the 2004 beam time, COMPASS was supplied with a tertiary polarized
$\mu^{+}$-beam at 160\,GeV/c with a nominal intensity of
$40\cdot10^6$\,s$^{-1}$. The momenta of individual beam particles were
measured with scintillator hodoscopes surrounding three beam-line dipole
magnets (Beam Momentum Station). The target consisted of two oppositely
polarized cells of $^6$LiD granulate in a liquid helium bath. Each
cell had a length of 60\,cm and a diameter of 3\,cm. About 30\% of the
2004 data set have been reprocessed with a newer version of the event
reconstruction software CORAL, in which a small inefficiency in the
reconstruction code of the beam momentum measurement system has been
cured. Only this portion of the data set is considered for the
luminosity analysis.\\ Due to an accelerator problem, COMPASS received
only half of its nominal beam intensity for half of the data taking
time under consideration here. Although this seems like an unpleasant
problem to deal with at first glance, it provides a valuable tool to
check the consistency of the obtained luminosity result. Although many
rate-dependent factors enter in the luminosity determination, cross
sections measured with the two different beam intensities have to be
identical. It is shown in Sec.\ \ref{sec:7} that this is in fact the
case.

\section{Data Selection}
\label{sec:4}\begin{figure}[t]
\centering
\begin{minipage}[t]{0.42\textwidth}
\centering
\includegraphics[width=\textwidth]{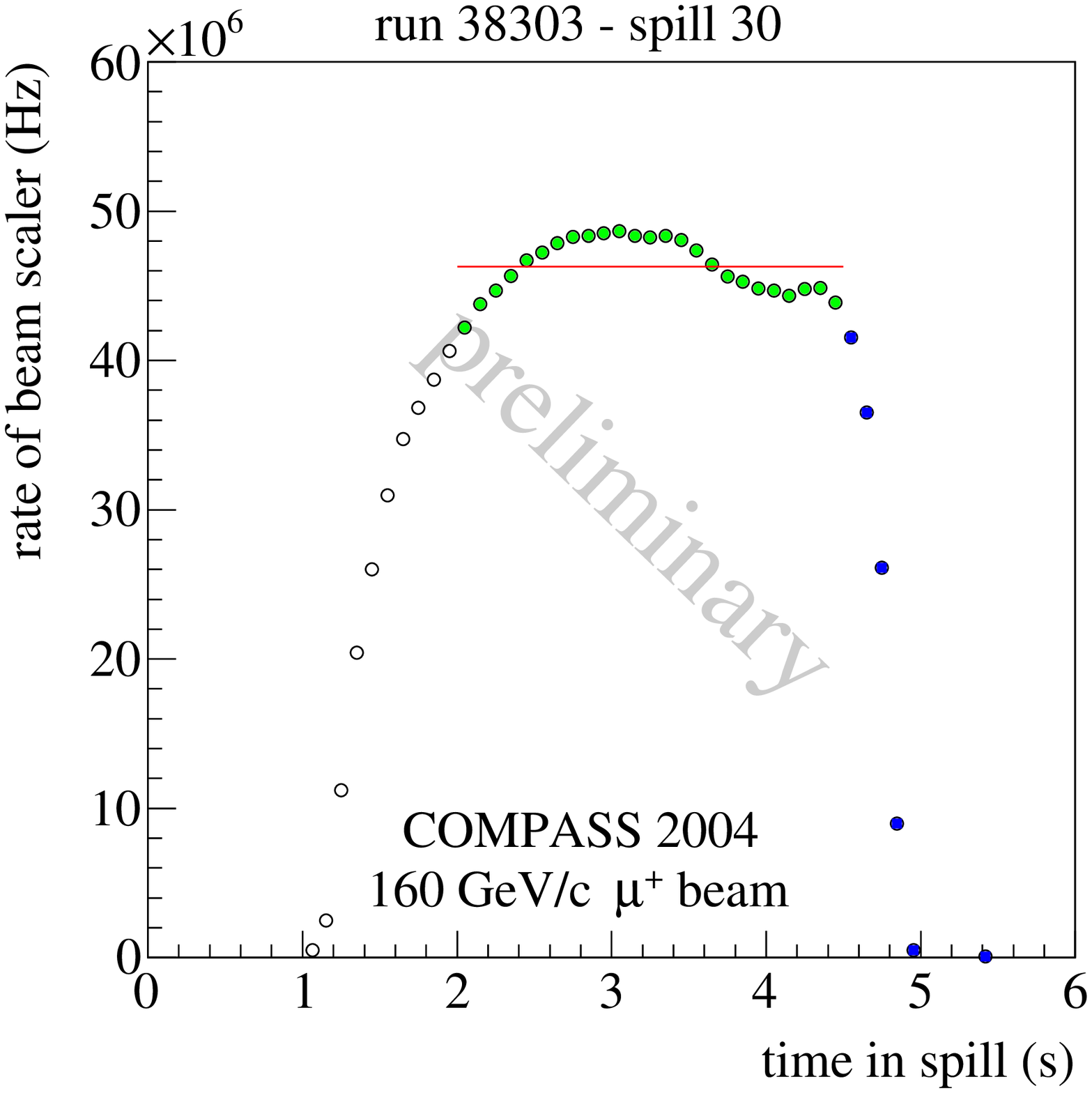}
\caption{Spill structure for spill 30 of run 38303. The green markers
  indicate time bins which are used for the analysis starting at
  $t_1=$ 2\,s. The blue markers indicate bins which are excluded by the $t_2$
  cut ($t_{30,2}=4.5$\,s). The red line indicates the average rate.}
\label{fig:spill1}
\end{minipage}
\hspace{0.03\textwidth}
\begin{minipage}[t]{0.42\textwidth}
\centering
\includegraphics[width=\textwidth]{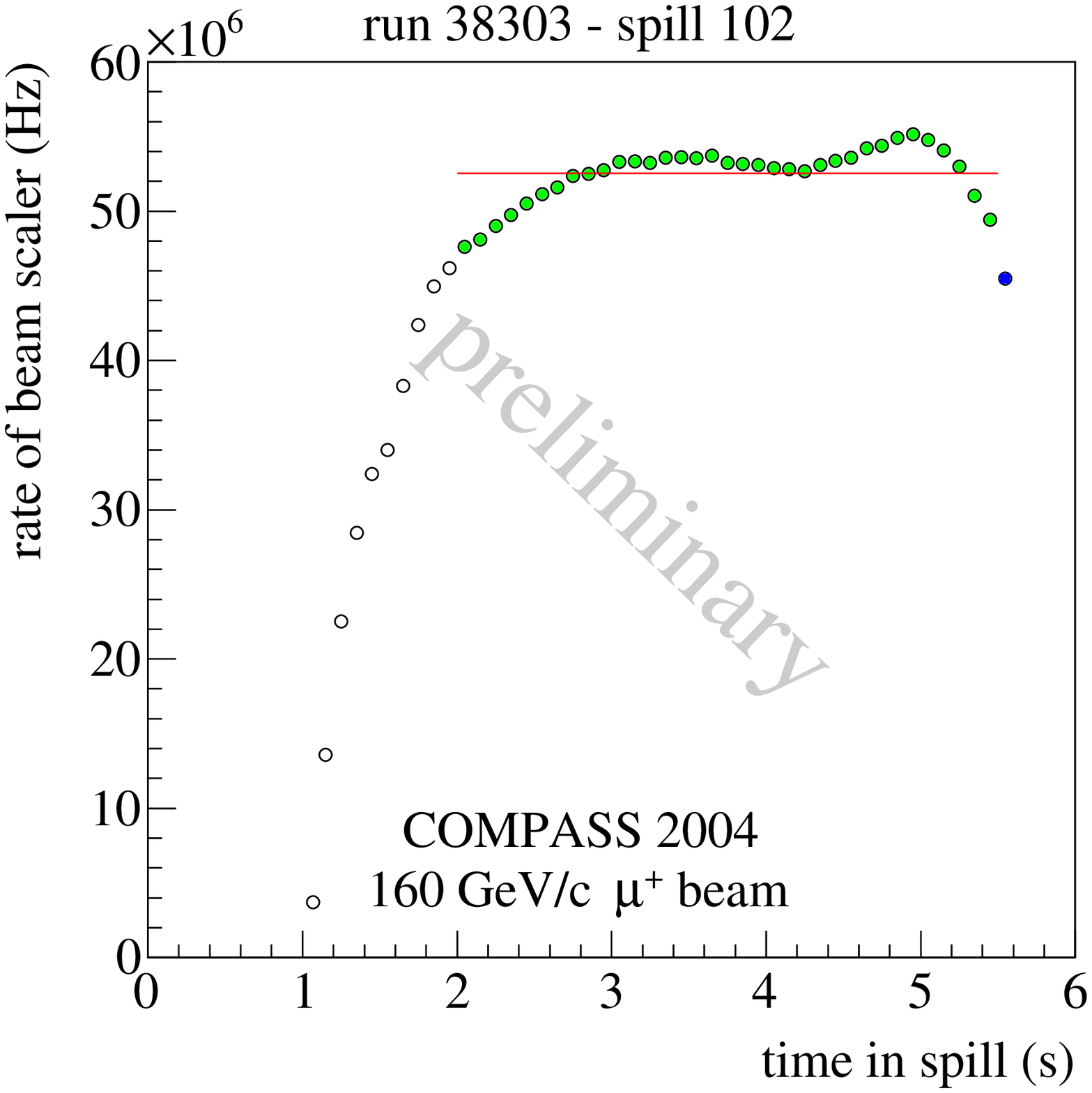} 
\caption{Spill structure for spill 102 of run 38303 ($t_{102,2}=5.5$\,s).}
\label{fig:spill2}
\end{minipage}
\end{figure}
After the beam intensity increases over the first second of
each spill, it is stable within $\sim 10$\%. It is known that
the COMPASS beam from the SPS can be poorly debunched in this period,
which makes the estimations of dead times very difficult because
beam and halo particles are not independent in their relative
timing anymore. As the corrections shall be applied as scaling factors
for each spill, only the flat top of the beam is selected for the
normalized analysis. The flat top of spills is defined to start at
time $t_{1}=2$\,s after the begin-of-spill signal from the SPS. The length of the flat top can vary between
different spills. Thus, the time in spill $i$ after which the data is
discarded from the analysis, $t_{i,2}$, is determined for each spill
individually. $t_{i,2}$ is defined as the time when the instantaneous beam
rate in the spill has dropped below 90\% of the average beam rate in
the spill.  The rate of incident beam particles is measured by a
scaler which counts the number of signals from a scintillating fiber
detector which is located just in front of the target (FI02Y).
Figures \ref{fig:spill1} and \ref{fig:spill2} show two
examples of spills with different average intensities and lengths.\\
The removal of spills which are affected by detector or data
processing problems is essential for the determination of a correct
luminosity for cross section measurements. Spills are
removed from the data sample if they fall below their neighboring
spills in one of the following four figures of merit:
\begin{itemize}
\item Number of primary vertices per reconstructed event.
\item  Number of outgoing tracks in the primary vertex per reconstructed event.
\item Number of beam tracks per reconstructed event.
\item Ratio of the number of reconstructed events in the spill over
  the number of triggered events during data taking. This criterion removes spills which are affected
  by very rarely occurring crashes of the reconstruction software or losses of
  data files due to tape problems.
\end{itemize}
The first three criteria are also applied in all COMPASS
spin-asymmetry analyses, whereas the last criterion is just needed for
absolutely normalized analyses.
After these strict quality cuts, 54624 of 73591 spills are
retained and are further used for the luminosity determination and
cross section analyses.

\section{Beam Flux Measurement}
\label{sec:5}
\begin{figure}[t]
\centering
\begin{minipage}[t]{.85\textwidth}
\centering
\resizebox{0.5\textwidth}{!}{
\includegraphics{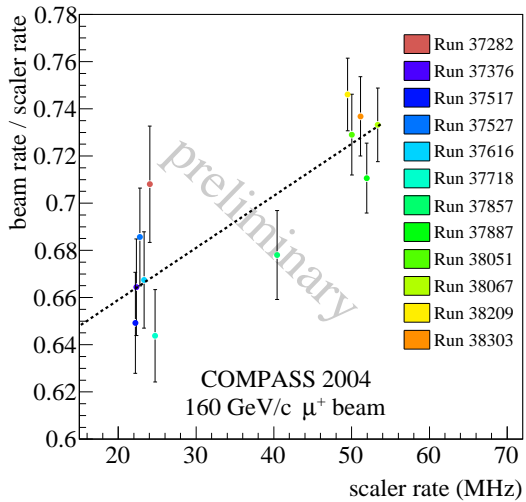}}
\caption{Calibration factor between $R_{\text{beam}}$ and
  $R_{\text{Sc}}$ as a function of $R_{\text{Sc}}$. The rate
  dependence is due to dead times in the scaler system. The dashed
  line indicates the function used for the spill-by-spill
  beam flux calibration.}
\label{fig:beam}
\end{minipage}
\end{figure}

The rate of particles measured with the scaler on FI02Y, $R_{\text{Sc}}$, does not equal the rate of beam particles
$R_{\text{beam}}$ from equation (\ref{eq:beam}) because not all beam particles which
cross the beam counter also cross the complete length of the fiducial
target volume. The
geometrical acceptance of the target w.r.t.\ the COMPASS muon beam is 65\%.
Furthermore, the rate measurement with the scaler system and the beam counter
can be affected by detection inefficiencies and dead times. A
calibration of $R_{\text{Sc}}$ with an unbiased measurement of
$R_{\text{beam}}$ is thus required. This calibration is performed on a
sub-sample of twelve runs by counting the number of reconstructed beam
tracks in random-trigger events\footnote{random triggers lead to a read-out of all detector electronics and are completely uncorrelated to the presence of beam tracks or scattering events.}.
The rate of beam particles measured in random-trigger events is
\begin{align}
R_{\text{beam}} = \frac{N_{\text{beam tracks}}}{\Delta t \cdot
  N_{\text{random triggers}}}
\end{align}
with the time window $\Delta t = 3.8$\,ns in which the detectors in
the beam telescope are fully sensitive to traversing beam
particles. $N_{\text{beam tracks}}$ counts the beam particles which
are retained after the fiducial target cut and the requirement of a
beam momentum measurement. The efficiency of the beam momentum
reconstruction is 93\% \cite{COMPASS}. The small inefficiency is automatically included in
the beam rate measurement from random-trigger events so that the
resulting luminosity will correctly contain only the portion of
the beam which is usable for the measurement of the yield of final
states $\tilde{N}$ in equation (\ref{eq:xs}). About every sixth random trigger contains a
reconstructed beam track at nominal intensity. 
The resulting calibration is shown in Fig.\ \ref{fig:beam}. The rate
dependence of the calibration constants is due to dead times in the
scaler system. The difference of the calibration between the half and
full intensity runs of more than 10\% proves to be correct in the
comparison of the inclusive muon scattering cross section in Sec.\
\ref{sec:7}. The systematic error on this calibration factor is
conservatively estimated to be 5\%, which is given by the RMS of the
twelve data points.\\ The rate of recorded random triggers, which was
about $100/(t_{i,2}-t_1)$ per spill in the 2004 beam time, will be
increased for future measurements for which a good luminosity
normalization is required, e.g.\ the DVCS measurement. The beam flux
will be estimated with negligible systematic uncertainty, if several thousand
reconstructed beam tracks per spill are recorded in random-trigger
events.

\section{Corrections for Dead Times}
\label{sec:6}
The DAQ dead time $d_{i,\text{DAQ}} $ from equation
(\ref{eq:effLumi}) is defined as the fraction of data taking time in
spill $i$ in
which no events can be recorded because the DAQ system is busy with
the acquisition of data from previously triggered events. It is
measured without any uncertainty in COMPASS by
counting the number of trigger attempts which were accepted by the DAQ
system and the total number of trigger attempts. The ratio of those
two numbers gives the life time $1-d_{i,\text{DAQ}}$ of the
system. It is measured in each spill in the flat top of the beam
($t\in [t_1,t_{i,2}]$, see Sec. \ref{sec:4}).
The trigger rate during the 2004 run at full intensity was
about 11\,kHz which resulted in a DAQ dead time of 9\%.

\begin{figure}[t]
\centering
\begin{minipage}[t]{0.42\textwidth}
\centering
\resizebox{\textwidth}{!}{%
\includegraphics{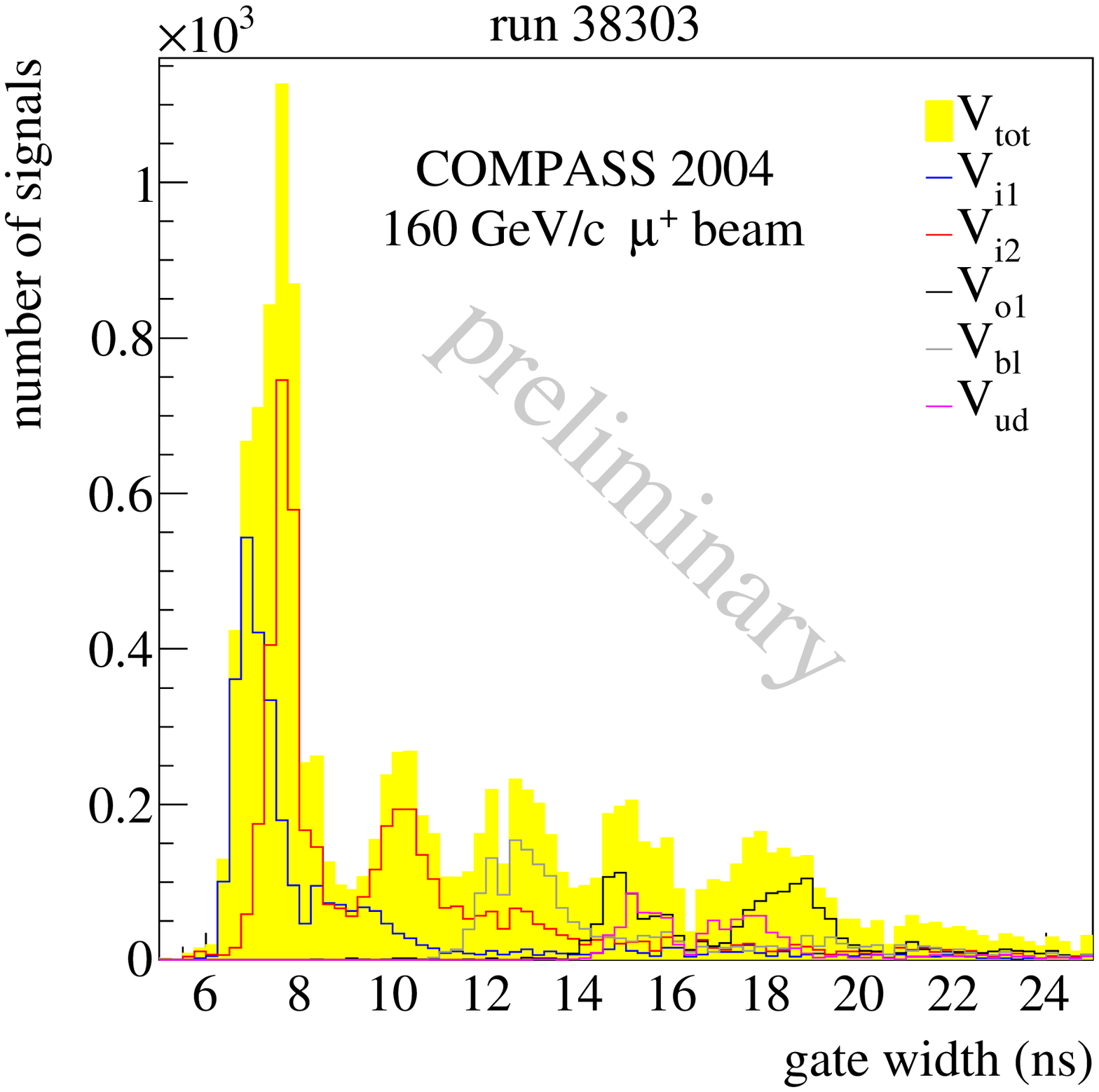}}
\caption{Gate width distribution of the veto signals. The signal
  $V_{\text{tot}}$ is made from a logical or of the individual veto
  signals $V_{\text{i1}}$ to $V_{\text{ud}}$. The veto counters close
  to the beam have shorter gates, but fire with higher rates than the
  more remote counters.}
\label{fig:vetoGates}
\end{minipage}
\hspace{0.03\textwidth}
\begin{minipage}[t]{0.42\textwidth}
\centering
\resizebox{\textwidth}{!}{%
\includegraphics{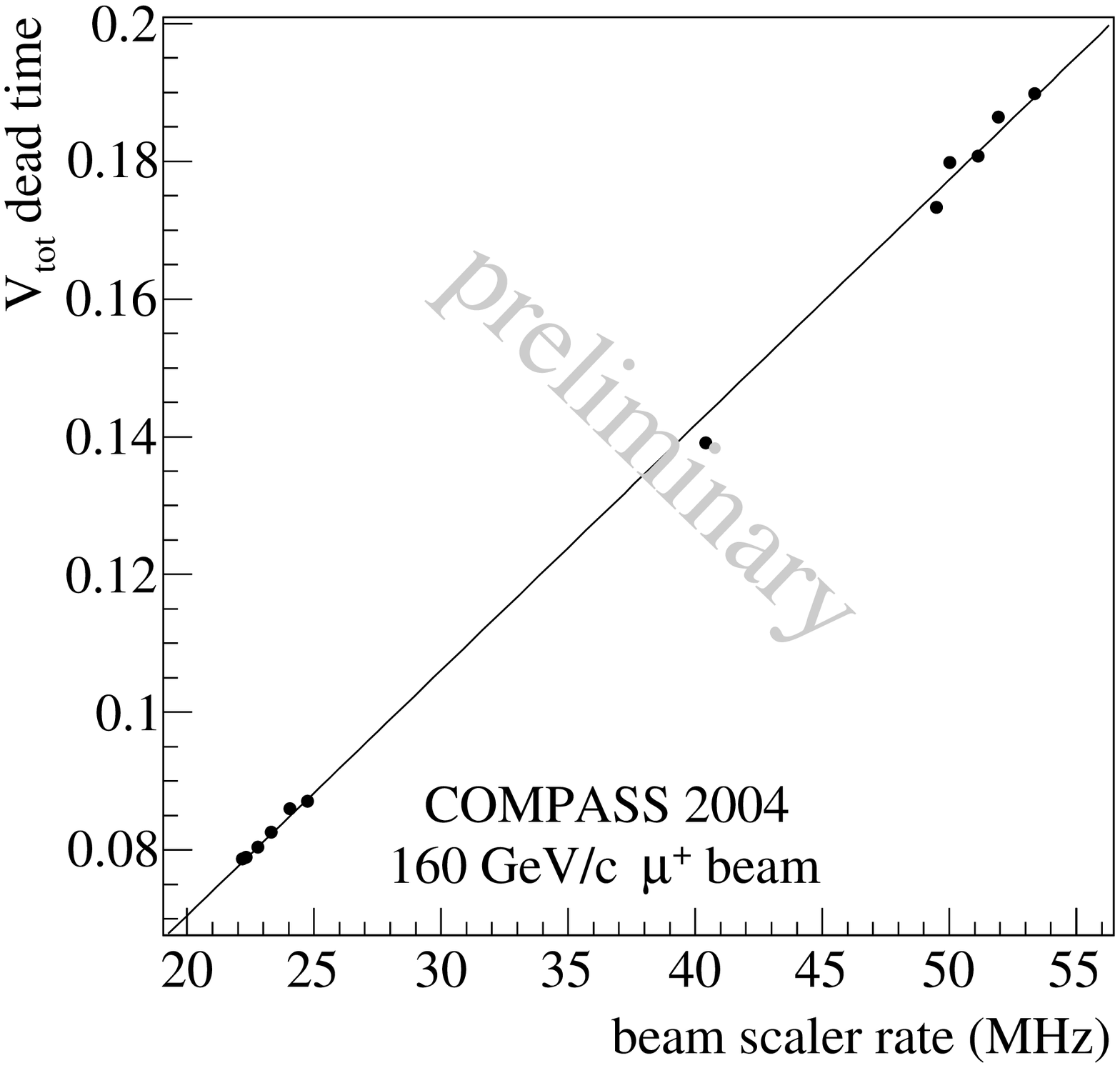} }
\caption{Veto dead time as a function of the beam scaler rate. At full
intensity, the dead time reaches a maximum of 19\%.}
\label{fig:vetoDeadTime}
\end{minipage}
\end{figure}

The second source of dead times in the COMPASS experiment is the veto
system of the muon triggers. These consist of
coincidences between scintillator-hodoscope pairs with target pointing for
different scattering kinematics. The hodoscopes are shielded by iron and concrete absorbers to ensure muon identification. The veto counters surround the beam
region upstream of the target to ensure that the trigger coincidence
was not due to one of the numerous halo tracks of the muon beam,
for details please be referred to
\cite{triggerNIM}. The veto dead time is defined as the fraction of data
taking time during which no triggers can be accepted because veto
signals are present, i.e.\ the duty cycle of the veto signal. The
number of veto pulses in the flat top of each spill is counted with a
scaler system. In combination with the gate width distribution of the
individual signals contributing to the veto system, which is
shown for one run in Fig.\ \ref{fig:vetoGates}, it allows the calculation of the
duty cycle. Figure \ref{fig:vetoDeadTime} presents the resulting dead
times for the same twelve runs that were used for the beam flux
estimation. The linear function fitted to the data is used to
estimate the veto dead time for all spills used in the
analysis. The absolute systematic error of the veto dead time is estimated to
be 0.03, which arises from the maximal difference with other measurements of the veto dead time during
data taking \cite{Pretz}.
\section{Luminosity Results and Determination of the Structure
  Function $F_2$}
\label{sec:7}
The COMPASS target in the 2004 data taking consisted (by number of
nucleons) of 42.3\% deuterium, 42.5\% lithium, and 15.2\% helium.  The
number of nucleons per unit area was $3.44\cdot 10^{25}$\,cm$^{-2}$
\cite{target} which was measured with a relative systematic
uncertainty of 2\%. The relative systematic uncertainty of the beam
flux calibration is 5\%. The reconstruction efficiency has a relative
systematic error of 1.8\%, and the veto dead time is measured with an
absolute uncertainty of 0.03. From these individual contributions, the
overall systematic
uncertainty is conservatively estimated to be 10\%.\\
The effective integrated luminosity for the discussed data sample,
corrected for the DAQ dead time, is $\tilde{L}=142.4$\,pb$^{-1}$. The
correction for the veto dead time reduces this number to
122.6\,pb$^{-1}$. Please note that not all triggers in COMPASS include
the full veto system. The triggers for the quasi-real photo-production
regime ($Q^2<$0.5\,(GeV/c)$^2$, where $Q^2$ is the negative four
momentum transfer of the muon), which are used for the
soon to-be-released high-$p_T$ hadron production cross section, are
subject to a much lower veto dead time. The so-called inclusive middle
trigger is used for the determination of the cross section for
inclusive muon scattering and the subsequent extraction of the
structure function $F_2$ in the deeply-inelastic scattering regime at
$Q^2>$1\,(GeV/c)$^2$. The scintillator hodoscopes contributing to this
trigger are fully efficient. Events are accepted if the beam energy
is $E_{\mu}\in[140,180]$\,GeV and the relative energy loss of the muon
$y=(E_{\mu}-E_{\mu '})/E_{\mu}$ is greater than $0.1$. Furthermore, the
extrapolated beam track is required to cross the full length of the
fiducial target volume (same cut as in the beam flux determination of
Sec.\ \ref{sec:5}) and the muon scattering vertex position must lie in
the fiducial target volume.

\begin{figure}[p]
\centering
\begin{minipage}[t]{0.85\textwidth}
\centering
\resizebox{0.6\textwidth}{!}{
\includegraphics{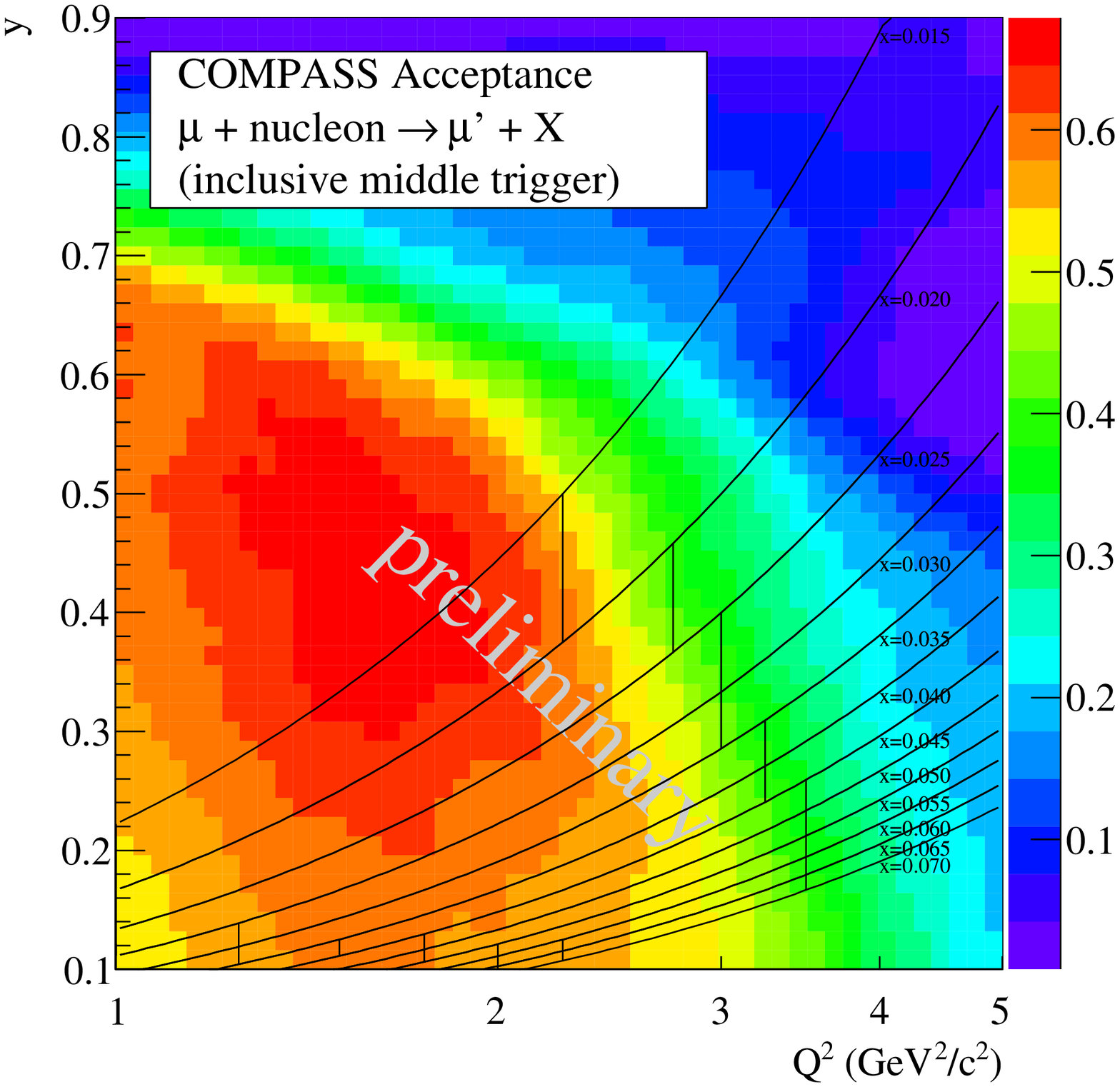} }
\caption{Acceptance for inclusive muon scattering in the inclusive
  middle trigger of COMPASS. It is determined with a Monte Carlo
  simulation consisting of LEPTO and GEANT3. The $x_{\text{Bj}}$-bins used
  in the analysis are indicated by the lines in the plot.}
\label{fig:acc}
\end{minipage}
\end{figure}
\begin{figure}[p]
\centering
\begin{minipage}[t]{0.85\textwidth}
\centering
\resizebox{\textwidth}{!}{
\includegraphics{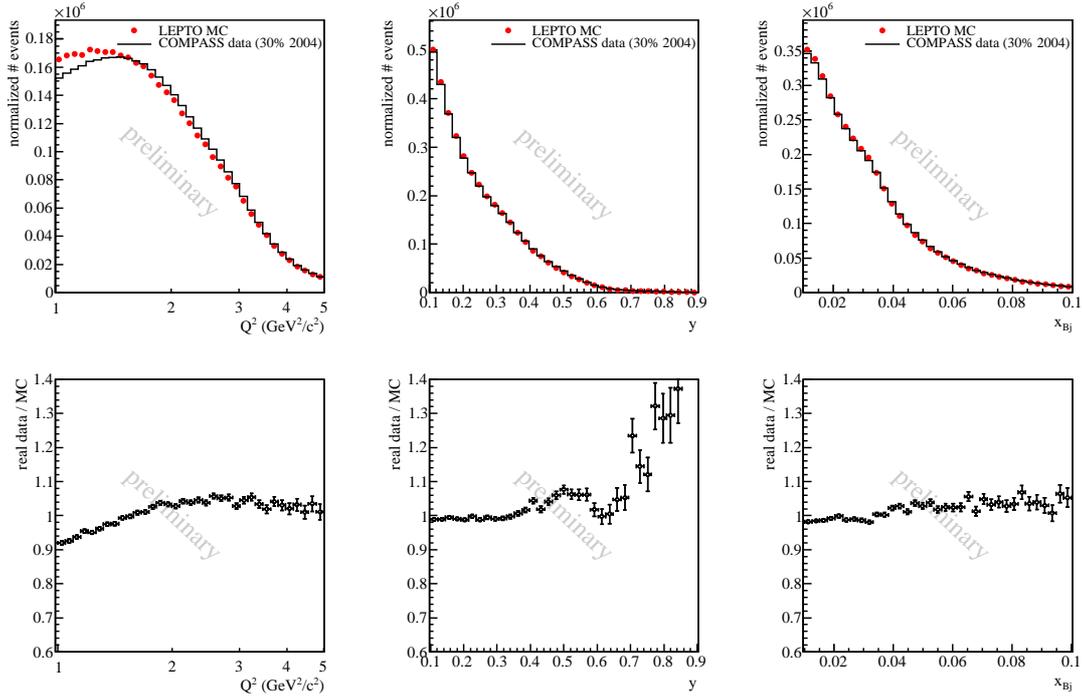} }
\caption{Comparison of real data (radiative correction applied) and Monte Carlo simulation in three
  kinematical variables. The slight deviation from unity might hint
  towards an incomplete description of the acceptance.}
\label{fig:RDoverMC}
\end{minipage}
\end{figure}
\begin{figure}[b]
\centering
\begin{minipage}[t]{0.85\textwidth}
\centering
\resizebox{0.575\textwidth}{!}{
\includegraphics{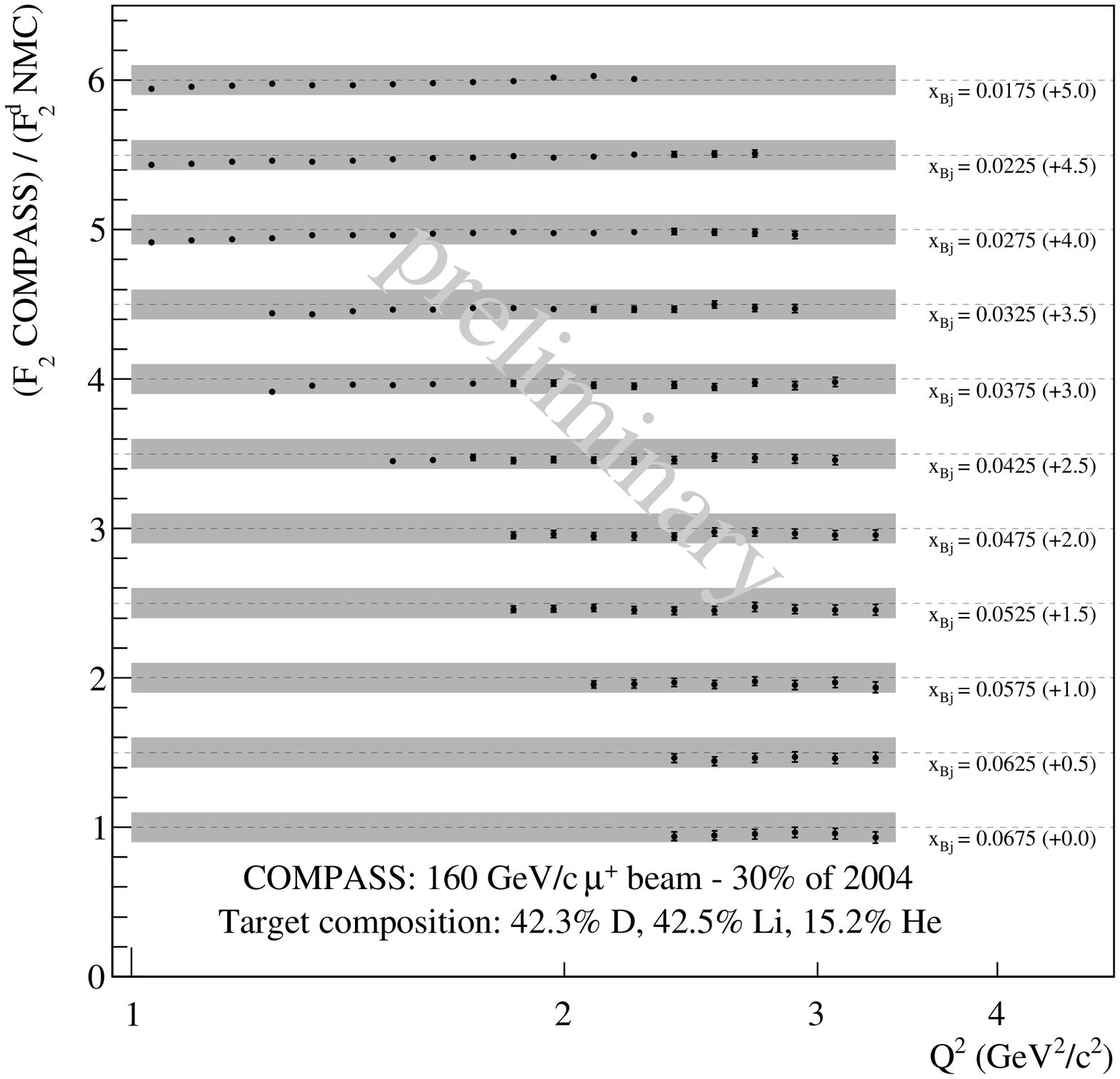} }
\caption{Comparison of the structure function $F_2$ extracted from
  COMPASS data with the parametrization from NMC \cite{NMC}. The ratios for the
  individual $x_{\text{Bj}}$ bins are offset by constant factors as
indicated on the right hand side of the plot. The gray bands indicate
the 10\% normalization uncertainty arising from the systematic error
on the luminosity. All value lie within the bands, thus proving
consistency with the NMC result. The slight deviations
from unity at the lower values of $Q^2$ can be due to  wrong descriptions
of the acceptance and to the fact that the radiative corrections have
not been iterated.}
\label{fig:F2ratio}
\centering
\resizebox{0.575\textwidth}{!}{
\includegraphics{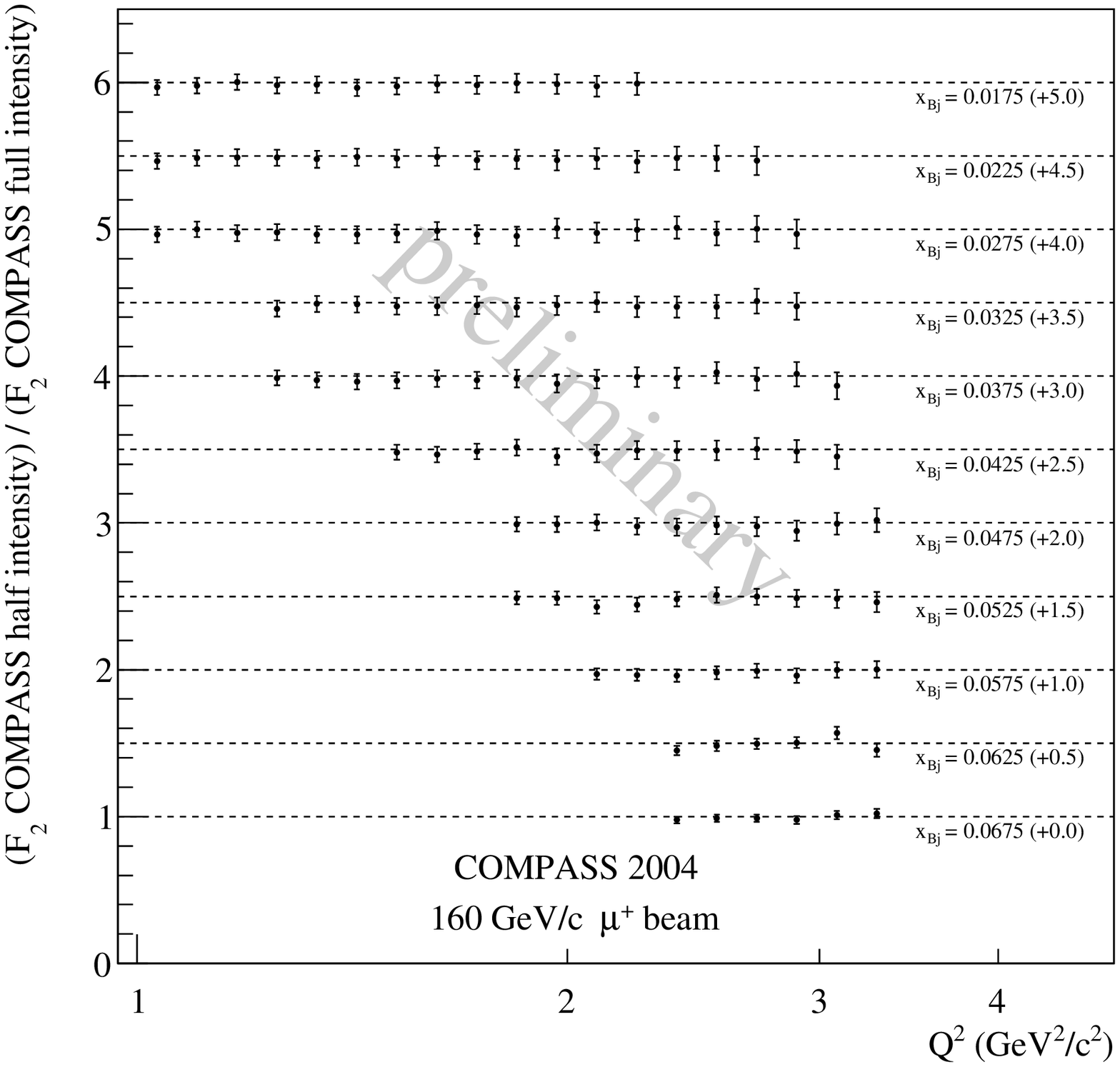} }
\caption{Comparison of $F_2$ determined from COMPASS with half and
  full beam intensities. The ratios for the
  individual $x_{\text{Bj}}$ bins are offset by constant factors as
indicated on the right hand side of the plot. Despite the many rate dependent factors that enter into the
  luminosity estimation, the results are fully consistent with each
  other.}
\label{fig:F2period}
\end{minipage}
\end{figure}

The acceptance factor $\epsilon=\epsilon(Q^2,y)$ from equation
(\ref{eq:xs}) is applied as a weighting factor on an event-by-event
basis. It has been determined with a Monte Carlo simulation of
inclusive muon scattering in the COMPASS experiment. The event
generator LEPTO \cite{LEPTO} has been used with the parton
distribution functions from MSTW 2008 \cite{MSTW}, including
$F_{\text{L}}$. The generated events have been transported through the
spectrometer with a GEANT3 \cite{GEANT} based program and analyzed
with the same reconstruction software (CORAL) which is used for
processing real experimental data. The acceptance as a function of
$Q^2$ and $y$ is shown in Fig.\ \ref{fig:acc}. It reaches a maximum of
66\% which is mostly due to the partial phase space coverage of the
selected muon trigger hodoscopes. A comparison of kinematical distributions of
reconstructed Monte Carlo events and real data events, as shown in
Fig.\ \ref{fig:RDoverMC}, indicates slight
disagreements of up to 10\% in some kinematical regions. This points
towards an incomplete description of the acceptance of the same
order of magnitude. Since it is the sole purpose of this analysis to
check whether the normalization of the luminosity is correct within
the systematic uncertainty of 10\%, this disagreement has not been
further investigated.\\ The structure function $F_{2}$ of the nucleon
is given by the cross section for inclusive scattering with
kinematical factors in the following way:
\begin{align}\label{eq:F2}
F_2(x_{\text{Bj}},Q^2) =& \frac{d^2\sigma_{1\gamma}(x_{\text{Bj}},Q^2,E_{\mu})}{dx_{\text{Bj}} dQ^2} \frac{x_{\text{Bj}}\cdot
Q^4}{4\pi \alpha^2} \\ \cdot& \{1-y(x_{\text{Bj}},Q^2,E_{\mu})-\frac{Q^2}{4 E_{\mu}^2} + (1-\frac{2 m^2}{Q^2}\cdot \frac{y(x_{\text{Bj}},Q^2,E_{\mu})^2+Q^2/E_{\mu}^2}{2[1+R(x_{\text{Bj}},Q^2)]})\}^{-1}
\end{align}
with the one-photon exchange (Born) cross section $\sigma_{1\gamma}$, the muon mass $m$, the Bjorken scaling variable $x_{\text{Bj}}$, the fine structure constant $\alpha$, and
the ratio of the longitudinal and transverse virtual photon
absorption cross sections\footnote{the same parametrization which had
  been used for the NMC $F_2$ extraction \cite{NMCcorr} was applied.}
$R(x_{\text{Bj}},Q^2)$. Since the experiments
use different targets and do not measure the Born cross section directly, so-called radiative
corrections have to be applied to the data. For a detailed discussion,
please see \cite{radiative}. The radiative event weight in a bin $(x_{\text{Bj}},Q^2)$ is
defined as
\begin{align*}
\eta (x_{\text{Bj}},Q^2)=\frac{\sigma_{1\gamma}(x_{\text{Bj}},Q^2)}{\sigma_{\text{measured}}(x_{\text{Bj}},Q^2)}.
\end{align*}
The values are taken from analyses in the NMC experiment \cite{NMCcorr}. The corrections are $<15$\% in the selected kinematical range ($y<0.5$). The fact
that these corrections have not been iterated with the new COMPASS
measurement is another reason why the extraction of $F_2$ presented
here is not to be regarded as a new precision measurement. Nuclear
effects on $F_2$ are negligible in the selected kinematical range. The
measured structure function can thus be directly compared to values of
the structure
function of the deuteron $F_2^{\text{d}}$ which are taken from a parametrization from the
NMC experiment which covers the complete kinematical reach of the
presented data. The ratio of the COMPASS result and the NMC
parametrization is presented in Fig.\ \ref{fig:F2ratio} for eleven bins
of $x_{\text{Bj}}$. The gray bands indicate the normalization
uncertainty of 10\% from the luminosity determination. The ratios lie
within the bands thus proving consistency with the NMC result.\\ Half of the presented data set was taken with 50\%
of the nominal beam intensity due to an accelerator problem in the
2004 beam time. Many rate dependent factors enter in the luminosity
determination. A comparison of $F_2$ determined from the two data sets
with different beam intensities provides an important consistency
check for the correctness of the result. Figure \ref{fig:F2period}
shows this comparison in which no deviations from unity are visible.

\section{Conclusions and Outlook}
\label{sec:8}
A method to determine the luminosity for data from the COMPASS experiment, taking into
account corrections for all dead times and inefficiencies, has been developed.
The luminosity for about 30\% of the 2004 muon-deuteron scattering data
set has been determined with a systematic error of
10\%. The structure function $F_2$ has been determined via the cross
section for inclusive scattering. A comparison of the result with a parametrization of
$F_2$ from the NMC experiment proves that the normalization is correct
within this uncertainty. The data set can hence be used for
the determination of unknown cross sections with an effective integrated luminosity of 142.2\,pb$^{-1}$. The COMPASS experiment
will soon publish the unpolarized cross section for the quasi-real
photo-production of charged hadrons with high transverse momenta. This
result will provide an important benchmark for the applicability of
perturbative QCD calculations which are and will be used for accessing
many aspects on the (spin-)structure of the nucleon in COMPASS.
Future measurements of other cross sections, for instance in the DVCS
program of COMPASS-II, will also be based on the techniques presented
here. Improvements on the beam-flux measurement with a higher
random-trigger rate and better measurements of the veto dead time will
significantly reduce the systematic uncertainty of the luminosity
measurement.
%

%

%

\end{document}